\input epsf
\documentstyle[pre,aps]{revtex}
\begin{document}
\draft
\title{The Force-Velocity Relation for Growing Biopolymers}
\author{A. E. Carlsson}
\address{Department of Physics\\
Washington University\\
St. Louis, Missouri 63130-4899}
\twocolumn[
\date{Submitted to Physical Review,\today}
\maketitle
\widetext
\begin{abstract}
The process of force generation by the growth of biopolymers
is simulated via a Langevin-dynamics approach. The interaction
forces are taken to have simple forms that favor the
growth of straight fibers from solution. The force-velocity relation
is obtained from the simulations for two versions of the 
monomer-monomer force field. It is found that the growth rate
drops off more rapidly with applied force than expected from
the simplest theories based on thermal motion of the obstacle. 
The discrepancies amount to a factor of three or more when
the applied force exceeds $2.5~kT/a$, where $a$ is the step
size for the polymer growth.  These results are explained 
on the basis of restricted diffusion of monomers near the fiber tip. 
It is also found that the mobility of the obstacle has little
effect on the growth rate, over a broad range.
\end{abstract}
\vskip 0.5cm
\twocolumn
]
\narrowtext
%\bibliographystyle{prsty}
%\newpage

\section{Introduction}

The growth of biopolymers is a key ingredient in the crawling motion 
and internal transport processes of almost all eukaryotic cells. 
They crawl among each other and over substrates by motion of 
the cytoplasm into protrusions known as lamellipodia, filopodia, 
or microspikes according to their shapes. The forces driving the 
extension of these protusions are believed to comes from the growth
of a collection of fibers assembled from monomers of the protein actin. 
The actin fibers are approximately 7~nm in diameter.
With no opposing force, they can grow at velocities\cite{Pollard86} 
of over $1~\mu$m/sec at physiological actin 
concentrations\cite{Cooper91,Marchand95} of 10--50$\mu$M; 
the velocities of the cell protrusions are 
typically\cite{Argiro85,Felder90} in the range of 
$0.1~\mu$m/sec. Actin fiber growth also can power the 
motion of bacteria and viruses through the cell cytoplasm.
The velocities usually range from $0.02$ to $0.2~\mu$m/sec,
but velocities up to $1.5~\mu$m/sec have been observed.
As they move, they leave behind ``comet tails" made up of actin
fibers\cite{Tilney89,Sechi97}. Recent experiments have studied the 
minimal ingredients necessary for such propulsion.
For example, Ref.~\onlinecite{Cameron99} shows that polystyrene beads 
coated with a catalytic agent for actin polymerization spontaneously
move in cell extracts at velocities of $0.01$ to $0.05~\mu$m/sec, 
forming comet tails similar to those caused by bacteria and viruses. 
It has also been shown recently that {\it Listeria} and {\it Shigella} 
bacteria can move in a medium much simpler than a cell extract, 
containing in addition to actin monomers only the proteins Arp2/3 
complex, actin depolymerizing factor, and capping protein. In particular,
myosin-type motors are not necessary for motion driven by actin 
polymerization.  The minimal ingredients lead to
velocities of $0.01$ to $0.02~\mu$m/sec; supplementation
of this mix with other ingredients including profilin, 
$\alpha$-actinin, and the VASP protein increases the velocities up
to $0.05~\mu$m/sec. To our knowledge, there have been
no measurements of the force-velocity relation for growing actin
filaments. However, recent measurements of the actin fiber 
density\cite{Abraham99} and Young's modulus\cite{Rotsch99}
at the leading edge of lamellipodia would suggest forces on
the order of 1~pN per fiber if all fibers are contributing
equally; this is roughly equal to the basic force unit for fiber
growth, $kT/a$, where $k$ is Boltzmann's constant, $T$ is
temperature, and $a = 2.7$~nm is the incremental fiber
length per added monomer.

Microtubules, which are thicker fibers (22~nm) assembled from tubulin 
subunits, also exert forces when they grow. Microtubule assembly and 
disassembly is crucial in intracellular processes such as mitosis, 
the formation of cilia and flagella, and the transport of nutrients
across the cell.  Recent measurements\cite{Dogterom97} 
on microtubules {\it in vitro} have yielded explicit force-velocity 
curves. At zero force, the velocity is about $0.02~\mu$m/sec; 
with increasing force, the velocity drops off roughly exponentially. 

It is clear that growth of the fiber against a force results
in a lowering of the system's free energy if the opposing force is
sufficiently small, since the exothermic contribution from
the attachment of monomers at the end of the polymer will outweigh
the work done to move the obstacle against the external force. 
The critical force at which polymerization stops is determined by 
the balance of these two contributions. However, it is not yet understood 
in detail what factors determine the rate of growth and the maximum 
force at which a useful speed can be obtained. The basic difficulty 
of the polymer's growth process is that when the  obstacle impinges 
directly on the fiber tip, there is not enough room for a new monomer 
to move in. Thus the rate of growth must be connected to the fluctuations 
of either the obstacle or the filament tip, which create temporary gaps 
between the tip and the obstacle. This effect has been treated explicitly 
in the ``thermal ratchet" model\cite{Peskin93}. In this model, 
one assumes that the obstacle must be a critical distance
$a$ from the tip for growth to occur. The fiber is assumed to be rigid. 
The growth rate is obtained by solution of a drift-diffusion type equation. 
For conditions of slow growth, i.e. in which the time to add a monomer 
is much longer than the time it takes the obstacle to diffuse a distance 
$a$, this equation can be solved analytically. The forward growth rate is
proportional to the probability that the obstacle-tip separation exceeds 
$a$. If depolymerization is sufficiently slow to be ignored, this yields 
the following dependence of the velocity $v$ on the opposing force $F$:
\begin{equation}
     v \propto \exp{(-Fa/kT)}
\label{exponential}
\end{equation}
where $k$ is Boltzmann's constant and $T$ is the temperature. 
This result is equivalent to application of the principle of detailed 
balance\cite{Hill87}, on the assumption that the depolymerization
rate is independent of the opposing force. This work has been
extended to flexible fibers at non-perpendicular 
incidence\cite{Mogilner96,Mogilner96a}, and to interacting systems of 
fibers\cite{Mogilner99}. For flexible fibers, it is again found
that the velocity is proportional to the probability forming of a gap
large enough to admit the next monomer. 

It is the purpose of this paper to evaluate the force-velocity
relation for growing fibers using a model more realistic than those
used previously. The model used to derive Eq.~(\ref{exponential})
does not explicitly treat the diffusion of monomers to the filament tip, 
but treats only the diffusive behavior of the variable describing 
the distance between the obstacle and the tip. It is assumed that 
once this distance exceeds $a$, that the monomers can enter with 
a fixed probability independent of the tip-obstacle distance. 
This assumption needs to be evaluated by explicit treatment of
the diffusion in the monomers. In addition, although the form of 
Eq.~(\ref{exponential}) is confirmed by the force-velocity relation 
for microtubules\cite{Dogterom97} the decay rate of the velocity 
with applied force was about twice as large as expected from 
Eq.~(\ref{exponential}).  One possible explanation of this,
suggested by Mogilner and Oster\cite{Mogilner99}, is subsidy effects 
between the thirteen fibers comprising a microtubule ``protofilament". 
We intend to investigate the extent to which other mechanisms 
can account for such discrepancies. 

\section{Model}

Our model system contains a fiber of protein monomers growing
perpendicular to a flat rigid obstacle in two dimensions. 
We will be mainly interested in the actin system, but the basic 
physics of our results is relevant to any fiber growing against 
an obstacle. Our choice of two dimensions is dictated mainly
by computational practicality: the simulations took over two weeks 
of CPU time on a Compaq 21264 processor and our preliminary studies 
indicate that the 
three-dimensional simulations take about 30 times longer.
The fundamental units of the simulation are the monomers; 
their internal and rotational degrees of freedom are assumed to be 
included in our effective interaction energies.  The motions of 
the monomers and the obstacle are treated via Langevin dynamics.
The $z$-direction is taken as the growth axis, with the obstacle
parallel to the $x$-direction.  The coordinates of the monomer 
centers-of-mass are given by $\vec r_i$, and the $z$-coordinate 
of the obstacle is called $Z$. The Langevin equations for this 
system are $\mu_i^{-1} d\vec r_i /dt=-\vec F_i+\vec f_i(t)$ for the
monomers and $\mu_O^{-1} d Z_O /dt=-F_O+f_O(t)$ for the obstacle, 
where the $\mu$'s are mobilities, $F$ denotes deterministic 
interaction forces, and $\vec f_i$ and and $f_O$ are random forces 
satisfying 
\begin{eqnarray}
\FL
      \langle f_i^x (t) f_j^x (t') \rangle 
  &=& \langle f_i^z (t) f_j^z (t') \rangle 
      = 2\mu_i^{-1} kT \delta_{ij} \delta (t-t'), \\
      \langle f_i^x (t) f_j^z (t') \rangle 
  &=& 0, \\
  {\rm and}~~
       \langle f_O (t) f_O (t') \rangle 
  &=& 2\mu_O^{-1} kT \delta (t-t') \quad . 
\end{eqnarray}

The Langevin equations are implemented with a finite time step $\Delta t$
following the procedure of Ref.~\onlinecite{Doi98}:

\begin{eqnarray}
\FL
      \vec r_i (t+\Delta t) 
  &=& \vec r_i (t) +\Delta t\mu_i \vec F_i (t) 
      + \vec g (t) \sqrt{kT \mu_i},\\
      {\rm and}~~
      Z (t+\Delta t) 
  &=& Z(t) + \Delta t\mu_O \vec F_O (t) + h(t) \sqrt{kT \mu_O}\quad ,
\end{eqnarray}
where $\vec g (t)$ and $h(t)$ are random functions with zero
time average, satisfying

\begin{equation}
      \langle g^x (t) g^x (t') \rangle  
     = \langle g^z (t) g^z (t') \rangle 
     = \langle h(t) h(t') \rangle 
     = 2 \Delta t \delta_{tt'}~~.
\end{equation}

To implement the last set of correlations, at each time
step we choose $g_x$, $g_z$, and $h$ from a uniform random
distribution random from $-\sqrt{6\Delta t}$ to $\sqrt{6\Delta t}$.

\subsection{Force Laws}

The obstacle experiences an external force of magnitude
$F_{\rm ext}$ in the $-z$ direction.

In the absence of reliable force fields for the monomer-monomer
interactions, we use a simple model form for the interactions which
has a linear filament as the lowest-energy structure. This form contains
two-body and three-body interactions. The two-body interactions
are repulsive and have the form
\begin{equation}
    V_2 (r_{ij}) = V^{\rm rep}\exp{[-\kappa_{\rm rep}(r_{ij}-a)]}\quad ;
\label{v2}
\end{equation}
the three-body interaction energy has the form
%\begin{equation}
\begin{eqnarray}
\FL
   V_3 (\vec r_{ij},\vec r_{ik}) 
 = && V^{\rm att} \exp{[-\kappa_{\rm att} (r_{ij}-a)]}\nonumber\\
   && \exp{[-\kappa_{\rm att} (r_{ik}-a)]} (\alpha + \cos{\theta_{ij}})~~~,
\label{v3}
%\end{equation}
\end{eqnarray}
with $0<\alpha <1$.  It is attractive for $\theta_{ij} >\cos^{-1} (-\alpha )$.
The monomer-obstacle interactions have only a two-body repulsive
component, and have the form
\begin{equation}
   W_2(z_i) = V^{\rm obst}\exp{(-\kappa_{\rm obst}|z_i-(Z-a)| )}\quad .
\label{w2}
\end{equation}
The forces are obtained as gradients of these energy terms.  
The energies are modified by subtraction of appropriate constants 
to force the interaction energy to go to zero at a cutoff distance 
$r_{\rm max}$ (in the case of the three-body terms this means that 
the energy vanishes if either $r_{ij}$ or $r_{ik}$ becomes greater 
than $r_{\rm max}$).

We use two parameter sets, whose values are given in 
Table~\ref{params}. These two parameter sets are chosen mainly
to sample different shapes of the ``basin of attraction" for
the addition of a monomer, and by no means exhaustively
sample the range of possible model force fields. 
The first corresponds to a narrow basin of attraction. 
The large value of $\alpha$ means that the three-body terms
are positive only for a small range of angles. This is partly
compensated for by the choice of prefactors to avoid the binding
energy becoming too small. We will call this the ``hard" force field. 
The corresponding energy contours are shown in Figure~1a. 
The width of the basin of attraction, or the region over
which the force pulls the next monomer into its minimum-energy
position, is about a tenth the size of a monomer, which
would correspond to a few {\AA} for actin monomers. 
Figure~1b shows the energy contours for the parameters corresponding 
to a wider basin of attraction, which is about a half the size of a 
monomer. We call this the ``soft" force field. For both of the force 
fields, the binding energies are very large compared to $kT$, 
so that monomer subtraction from the fiber does not occur in the 
simulations. This is a reasonable approximation; from the measured 
on and off constants in Ref.~\onlinecite{Pollard86}, the ratio of 
on to off rates at physiological actin monomer concentrations would 
be less than $0.01$.

With regard to the mobilities, the only physically relevant factor
is the ratio of the obstacle mobility to the monomer mobility, since
multiplicative changes in all the mobilities simply serve to scale
up the fiber growth velocities; these will thus factor out of our
velocity results, which are scaled by $1/\mu kT$. 
For most of our simulations, we use a mobility of the obstacle equal
to that of the monomers for simplicity. 
This would correspond to identifying the obstacle with a part
of a fluctuating membrane, rather than an entire rigid particle.  
We have varied the obstacle mobility  in a few cases, with results 
to be discussed below. 
%TAB1
\begin{table}
\tighten
\caption{Parameters used in simulations. Energies are given in units
of $kT$, $\kappa$-parameters in units of $a^{-1}$, and 
$r_{max}$ in units of $a$, the equilibrium monomer spacing.}
\begin{tabular}{lllllllll}
Force Field&$V_{\rm rep}$&$V_{\rm att}$&$V_{\rm obst}$
&$\kappa_{\rm rep}$&$\kappa_{\rm att}$&$\kappa_{\rm obst}$
&$\alpha$&$r_{\rm max}$\\
\tableline
Hard&141.3&6510&19.14&8.267&4.960&4.960&0.940&1.412\\
Soft&257.5&2151&27.44&7.666&4.600&4.600&0.770&1.522\\
\end{tabular}
\label{params}
\end{table}
%FIG1
\begin{figure}
\input psfig.sty
\centerline
{\psfig{figure=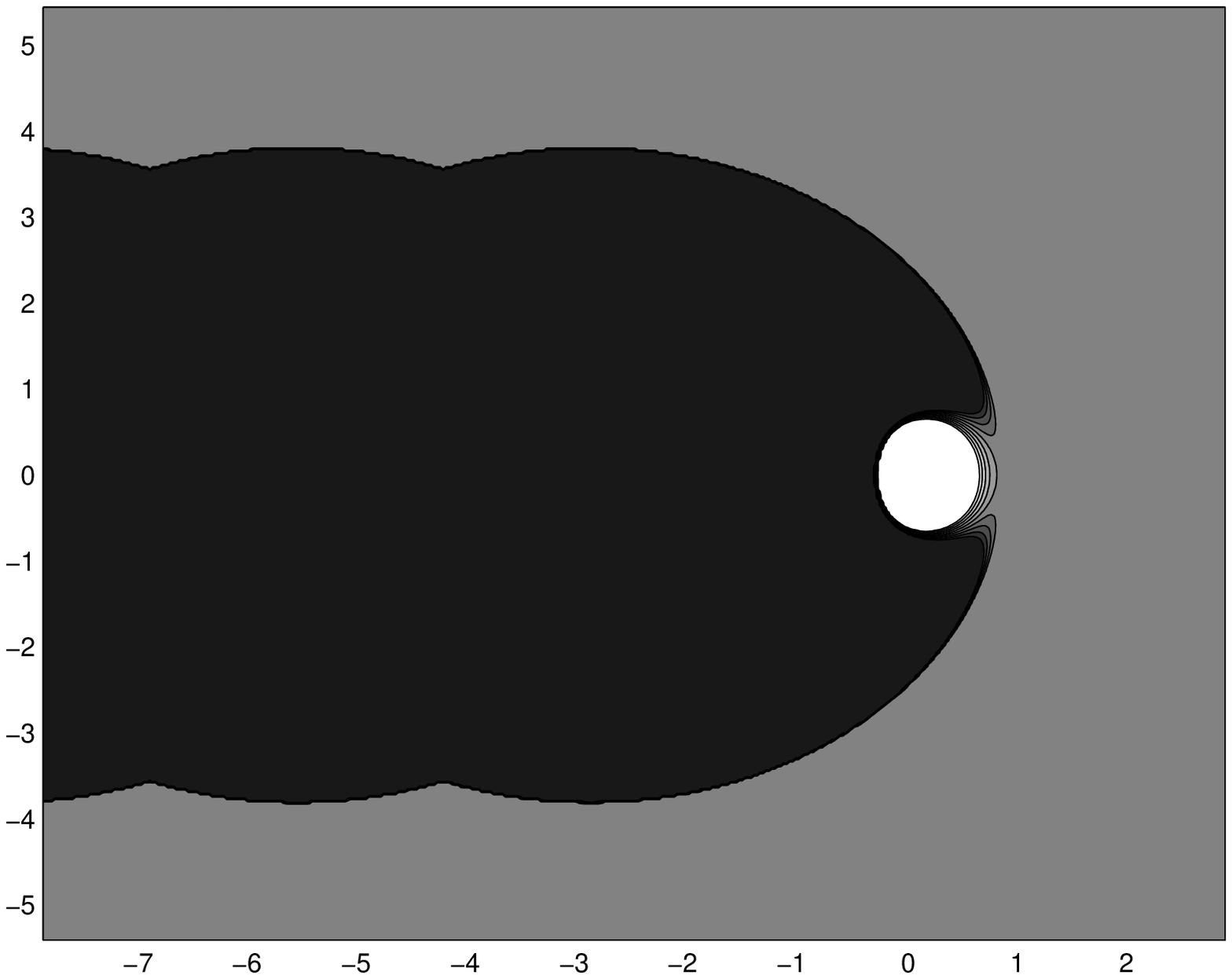,width=6cm,height=4.5cm}}
\input psfig.sty
\centerline
{\psfig{figure=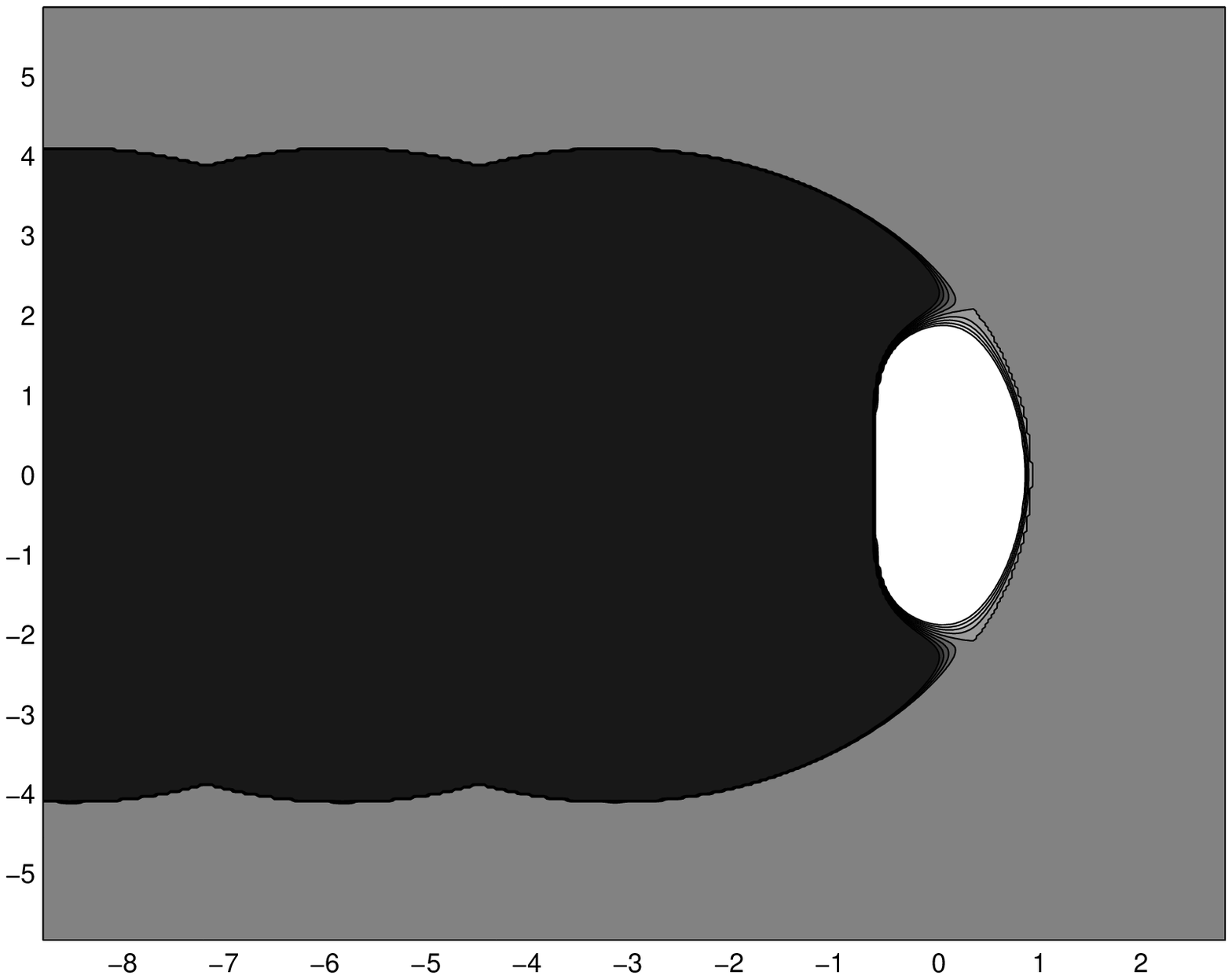,width=6cm,height=4.5cm}}
\caption{Energy contours for monomer approaching fiber tip
with hard (a) and soft (b) force fields. Contour heights
correspond to integer multiples of $kT$, with lighter corresponding
to lower energies. The length units are nm, assuming a monomer
step size of 2.7~nm as for actin.}
\end{figure}

\subsection{Filament-Growth Procedure}

A typical physiological concentration of actin ($10~\mu$M) is low 
in the sense that the average spacing between actin monomers is about 
60~nm, roughly 10 times the monomer size. This means that the 
probability that two free monomers are near enough to interact 
with each other is very small. 
For this reason we adopt a growth procedure in which only one free 
monomer at at time interacts with the tip. This is accomplished as 
follows. 
We start with a fiber of six monomers pointing in the 
$z$-direction, at their equilibrium spacing. A free monomer is then 
added at a point on a circle of radius $R$ centered on the next 
attachment site\cite{Note1} (defined as one monomer spacing beyond 
the monomer at the fiber tip). We choose $R=2.5a$, which places 
the added monomer well beyond the interaction range of the monomer 
at the tip. The relative probabilities of monomer addition at
different points on the circle are proportional to $\exp{[-W_2(z)/kT]}$. 
This weighting is accomplished by choosing a random number for each 
potential addition point; if this random number is less than 
$\exp{[-W_2 (z)/kT]}$, then this point is rejected and another
one is chosen. A new point is also chosen if the monomer overlaps
the fiber (i.e., its distance to the fiber is less than $r_{\rm max}$).
The system is then stepped forward in time according to the procedure 
described above, until one of two possible termination events occur:
\begin{itemize}
\item{1. The monomer diffuses outside of the $R$-circle.
In this case it is restarted on the circle as above.
If the obstacle abuts the fiber, the position of the monomer is
constrained to be out of the interaction range of the obstacle.}
\item{2. The monomer attaches to the tip. In this case,
another monomer is started on the $R$-circle.} 
\end{itemize}
In this way, the CPU time that is used in the simulation is
focused on the time that the monomers spend close to the tip. 
A typical snapshot of a simulation configuration is shown in Figure~2.
%FIG2
\begin{figure}
\input psfig.sty
\centerline
{\psfig{figure=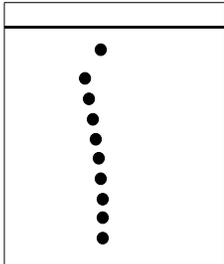,width=3cm,height=3.5cm}}
\caption{Typical fiber-obstacle configuration during 
simulations.}
\end{figure}

One can use the computed growth rates to predict growth rates for low  
concentrations $c$, by simply multiplying the computed rates by the 
probability $P(c)$ of finding a monomer inside the $R$-circle. 
We obtain this probability numerically as
\begin{equation}
   P(c)= \frac{1}{c}\int_{r<R} \exp{(-U(\vec r )/kT)} d^2 r~~,
\label{correction}
\end{equation}
where $U(\vec r )$ is the energy (from both fiber and obstacle) 
associated with placing a monomer at the point $\vec r$, 
and the coordinates are given with respect to the next attachment 
point\cite{Note2}. We plot our force-velocity relations in terms 
of the force acting between the obstacle and the fiber tip. 
This exceeds the external force applied on the obstacle by an amount 
corresponding to the viscous drag on the obstacle as it moves through 
the medium. The total force is thus given as $F=F_{\rm ext}+v/\mu_O$.

\section{Results}

%\subsection{Growth Statistics}
%
%Figure 3 shows a histogram for the growth velocities obtained in 2d 
%simulations where up to 30 monomers are added to the fiber tip,
%where we have used the ``soft" force field. 
%As expected, the distribution is Gaussian, with a width of roughly
%???. 
Our simulations involve 10 runs, each of which involves the addition 
of 30 monomers to the fiber tip. This corresponds to a statistical
uncertainty of $\sqrt{1/300}=6\%$ in the growth velocities.
Typical results for the fiber length as a function of time are
shown in Figure~3. Note that there are no backwards steps, because
the parameters that are used in the force field result in an
exothermic enthalpy for monomer addition that exceeds $kT$ by
at least a factor of 20. We use a concentration corresponding
to one monomer per square of side $20a$; in our model, velocities
at other concentrations would be given by a linear proportionality.
%FIG3
\begin{figure}
%\input psfig.sty
%\centerline
%\epsfxsize=175pt\epsfbox{Fig3.ps}
%\epsfxsize=155pt\epsfbox{Fig3.ps}
\epsfxsize=125pt\epsfbox{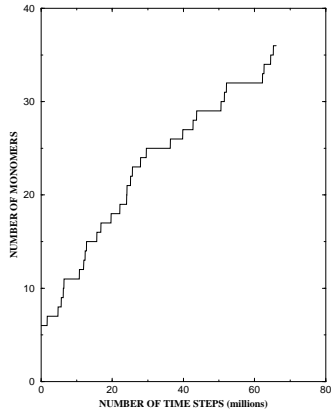}
%{\psfig{figure=Fig3.ps,width=3cm,height=3.5cm}}
\caption{Representative plot of number of monomers in fiber vs. number 
of time steps. Obtained for $Fa/kT=1.5$ and hard force field.}
\end{figure}

\subsection{Force-Velocity Relation}

Figure 4a shows growth velocity (solid circles) vs. applied force, for
the ``hard" force field ({\it cf.} Figure~1a).
For comparison, a curve proportional to $\exp{(-Fa/kT)}$ is shown. 
The simulation results give noticeably lower velocities 
at finite applied forces than the exponential prediction. 
The discrepancy is about $65\%$ at $Fa/kT=1$, and $85\%$ 
at $Fa/kT=2.5$. The results can be roughly fitted to different 
exponential curve, of the form $\exp{(-1.7~Fa/kT)}$.
Thus the growth velocity is much more sensitive to force than 
the thermal-ratchet model would predict. 
Figure~4b shows similar results for the soft force field 
({\it cf.} Figure~1b).  The free-fiber growth velocity is about 
twice that for the ``hard" force field, because the attraction basin 
is larger. The discrepancies between the simulation results and the 
analytic theory are comparable to those seen for the ``hard" force field,
but somewhat less pronounced. The discrepancy at $Fa/kT=2.5$ is
$70\%$, and the exponential fit curve is $\exp{(-1.5~Fa/kT)}$.
The open diamonds in Fig. 4b correspond to the results of varying
the mobility $\mu_O$; for the leftmost one the mobility is doubled,
and for the rightmost one it is reduced by a factor of ten. The
effects of these variations are very minor, as predicted by the 
``thermal-ratchet" model\cite{Peskin93}.
%%FIG4
\begin{figure}
%\input psfig.sty
%\centerline
%\epsfxsize=175pt\epsfbox{Fig4a.ps}
%\epsfxsize=155pt\epsfbox{Fig4a.ps}
\epsfxsize=125pt\epsfbox{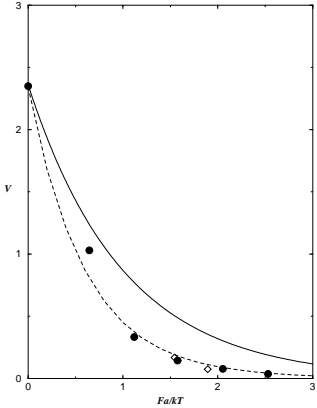}
%{\psfig{figure=Fig4a.ps,width=3.5cm,height=3cm}}
%\input psfig.sty
%\centerline
%\epsfxsize=175pt\epsfbox{Fig4b.ps}
%\epsfxsize=155pt\epsfbox{Fig4b.ps}
\epsfxsize=125pt\epsfbox{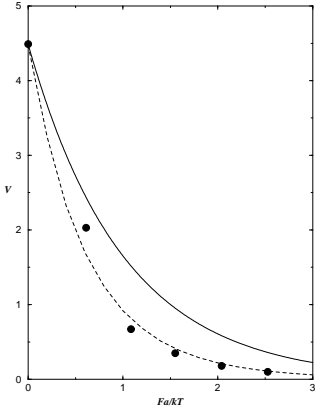}
%{\psfig{figure=Fig4b.ps,width=3.5cm,height=3cm}}
\caption{Growth rates (solid circles) for hard (a) and soft (b) force fields
vs. total force $F$. Rates given in units of $\mu kTc$, where
$\mu$ is the monomer mobility and $c$ is the concentration.
Force given in units of $a/kT$. Solid line corresponds to
exponential decay ({\it cf.} Eq.~(1)). Diamonds in (a) correspond to 
mobility enhanced by factor of 2 (left) and reduced by
factor of 10 (right). Dashed curves correspond to theory
of Eq.~(12).}
\end{figure}

%Figures 5a and 5b show similar results for three-dimensional simulations,
%although in this case the error bars are larger. For the soft force
%field, the discrepancies between the simulations and the Brownian
%ratchet theory are smaller than the error bars. However, in the 
%case of the hard force field, there is a discrepancy of  nearly
%a factor of two, still significantly greater than the error bars. 

\subsection{Interpretation}

We believe that the discrepancies seen in Fig.~4 result from the
restriction of monomer diffusion to the fiber tip by the impinging
obstacle. Such restriction will occur even when the obstacle is elevated
by a distance $a$ or more. Figure~5 shows energy contours for a 
monomer approaching the tip, when the obstacle  is elevated 
a distance $1.25a$ relative to its equilibrium position for
$Fa/kT=1.0$. The contours at at integer multiples of $kT$. 
The easily accessible paths corresponding to energies less than 
$kT$ are confined to a narrow band by the presence of the obstacle. 
This is expected to slow the diffusion to the tip. Effectively, 
the monomers must travel through a tunnel in order to get to the 
basin of attraction near the tip. Another possible explanation 
for the observed effect would be that even in the region with 
energy less than $kT$, there is a finite energy from the 
interaction with the obstacle. However, this energy is proportional 
to the the length scale of the interaction between the obstacle 
and the monomers. In a few cases, we have made this length scale 
five times smaller, and the velocities are unchanged to within a
few percent. Therefore, this monomer-obstacle interaction energy 
does not seem to be the major factor, but rather the blocking effects 
of the obstacle. 
%FIG5
\begin{figure}
\input psfig.sty
\centerline
%{\psfig{figure=Fig5.ps,width=4.5cm,height=6cm}}
{\psfig{figure=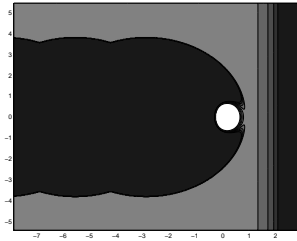,width=4cm,height=3.25cm}}
\caption{Energy contours for monomer approaching fiber tip
with hard force field, in presence of obstacle. 
Contours are as in Fig.~1.}
\end{figure}

To make this physical picture more precise, 
we have calculated the velocities for model fiber
configurations in which the obstacle is held at a fixed distance
from the fiber tip. The results are shown in Figs. 6a and b, for
the ``hard" and ``soft" force fields respectively. The distance $Z$ is
measured in units of the monomer size, and the edge of the obstacle
is defined as the point where the monomer-obstacle interaction energy 
is equal to $kT$.  Thus when $Z=0$, the interaction
energy of the last monomer in the fiber with the obstacle is $kT$.
In both cases, the velocity at $Z/a=1$ is nearly zero.  
Only for $Z/a > 2$ is the velocity within $20\%$ of the free-growth velocity. 

The appropriate generalization of Eq.~(\ref{exponential}) is then
the following:
\begin{equation}
   v(F) = \int^{\infty}_0 v(Z) P(Z,F) dZ~~~,
\label{vint}
\end{equation}
where $F$ is the applied force, $Z$ is the obstacle position,
and $P(Z,F)=(\rm const) \exp{(-E/kT)}$ is the 
probability of a certain value of $Z$. Here the obstacle-fiber
interaction energy is $E=W_2(z-Z)+FZ$, where $z$ is the z-coordinate 
of the last monomer in the fiber. 
Equation~(\ref{vint}) reduces to Eq.~(\ref{exponential}) if $v(Z)$ has the form
of a step function beginning at a $Z=a$, and
$W_2$ is sufficiently short-ranged. The dashed lines in 
Figs.~4a and 4b correspond to a numerical evaluation of
Eq.~(\ref{vint}). For both force fields, the agreement with
the simulation results is quite close, with only about
$20\%$ discrepancies occurring for small but non-zero forces. 
Thus the gradual rise of the velocity seen in Fig.~6, as opposed
to an abrupt jump, is at the heart of the observed effect.
%FIG6
\begin{figure}
%\input psfig.sty
%\centerline
%{\psfig{figure=Fig6a.ps,width=3cm,height=3.5cm}}
%\epsfxsize=155pt\epsfbox{Fig6a.ps}
\epsfxsize=125pt\epsfbox{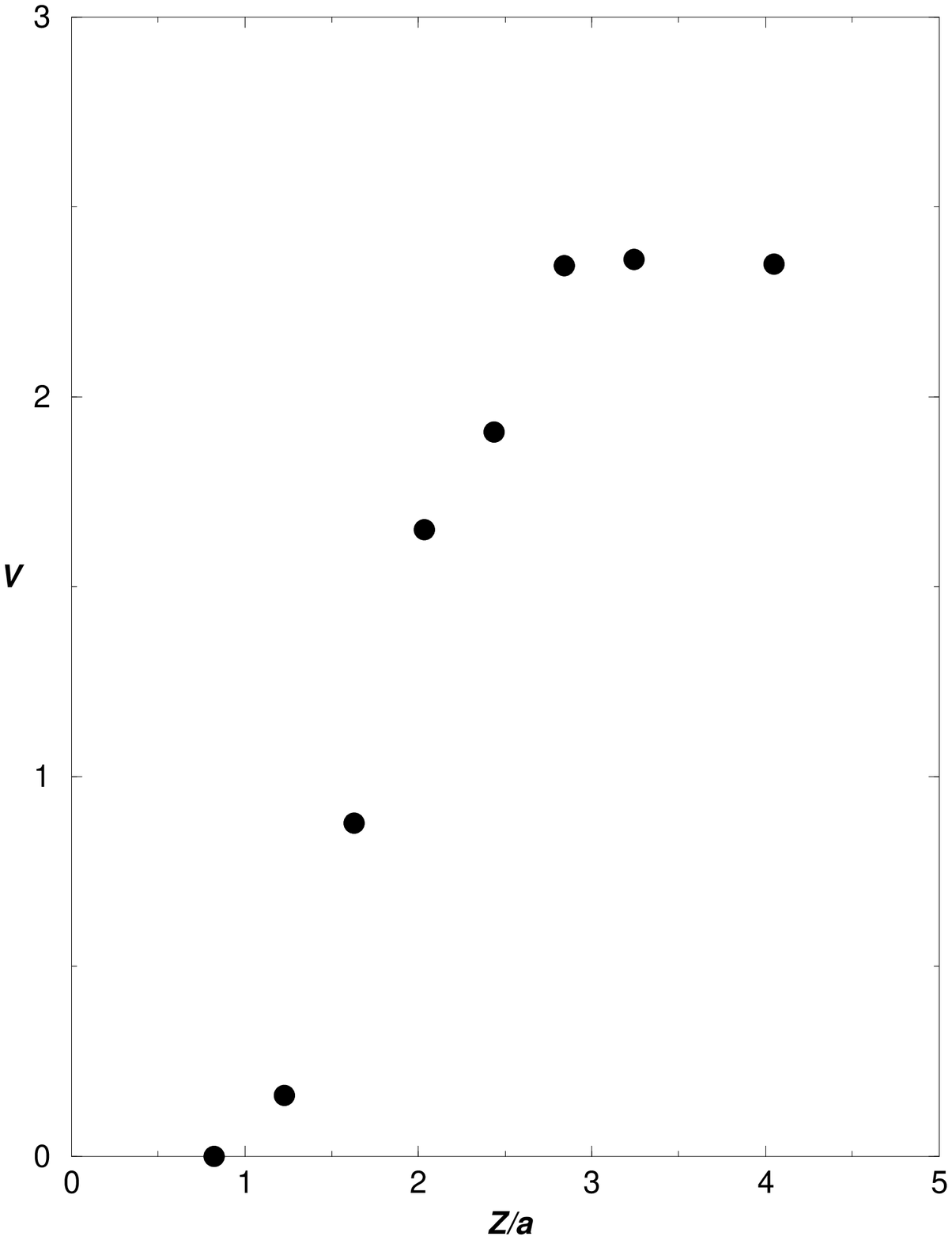}
%\input psfig.sty
%\centerline
%{\psfig{figure=Fig6b.ps,width=3cm,height=3.5cm}}
%\epsfxsize=155pt\epsfbox{Fig6b.ps}
\epsfxsize=125pt\epsfbox{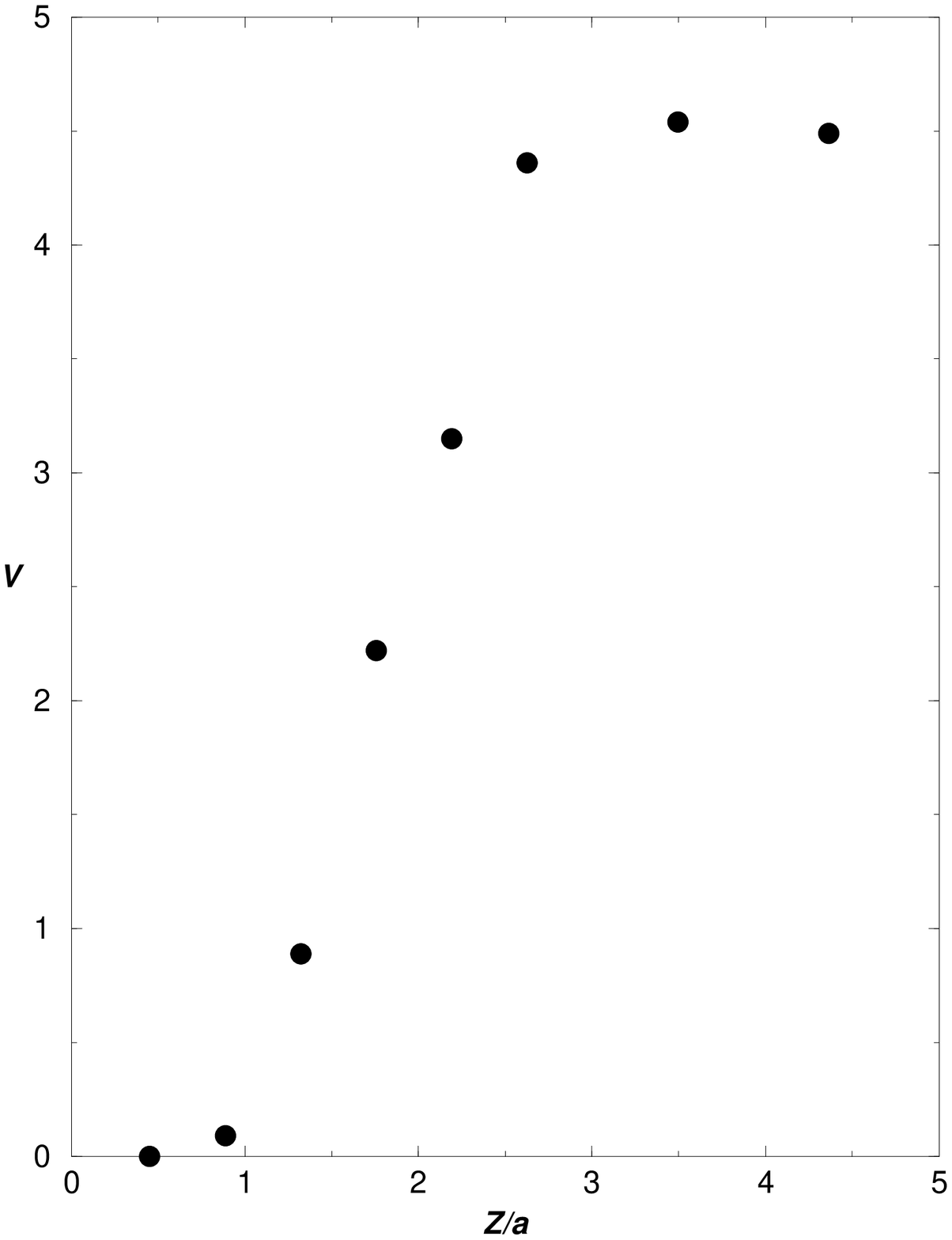}
\caption{Fiber growth velocity with fixed tip-fiber spacing, for
hard (a) and soft (b) force fields. $Z$ is measured relative to point at which
tip-fiber interaction energy is $kT$.}
\end{figure}

\section{Conclusions}

The physics underlying the above results is general enough 
that in most systems involving fiber growth against an obstacle,
one should expect a decay of velocity with applied force more
rapid than the simple exponential form (\ref{exponential}).
This may explain some of the discrepancies pointed out in connection
with the measured force-velocity relation of Ref.~\onlinecite{Dogterom97}. 
However, our application to microtubule growth is not quantitative
enough to determined whether the present effect exceeds the
subsidy effects discussed in Ref.~\onlinecite{Mogilner99}.
The results obtained here should be useful
in explaining the basic physics of motion based on actin
polymerization. For example, in a recent study van Oudenaarden
and Theriot\cite{Ouden99} have simulated the propulsion of 
plastic beads in cell extracts with a model based on a 
number of fibers exerting forces on the beads. In their simulations, 
an assumed form is taken for the probability of monomer addition 
to a fiber in terms of the time-averaged position of the fiber 
relative to the bead, or equivalently the force acting between the two. 
A better knowledge of the relationship between the force and the monomer 
addition rate can help pin down the validity of the assumptions 
underlying such simulations. Because the bead-motion
simulations include only the thermal energy required to achieve a
certain tip-obstacle spacing, it is likely that the addition rate
will drop off more rapidly with increasing force than is assumed
in Ref.~\onlinecite{Ouden99}.

The results obtained here are also expected to have noticeable
results on the structure of membranes that are being pushed forward
by collections of actin fibers. As a result of random fluctuations,
some fibers will eventually get ahead of others, and these will
be exerting larger forces on the membrane. If the velocity drops
off rapidly with the force, then these fibers will be slowed down
significantly. This will result in the membrane surface being
smoother than otherwise expected. Future work should treat such
many-fiber effects, and also explore the effects of fiber growth
angle and branching.

\acknowledgements

I am grateful to John Cooper for stimulating my interest in this 
project, and to Jonathan Katz and Elliot Elson for useful conversations. 
This research was supported by the National Institutes of Health 
under Grant Number GM38542-12.

%\bibliography{oster}

\begin{thebibliography}{10}

\bibitem{Pollard86}
T. Pollard, J. Cell Biol. {\bf 103},  2747  (1986).

\bibitem{Cooper91}
J.~A. Cooper, Ann. Rev. Physiol. {\bf 53},  585  (1991).

\bibitem{Marchand95}
J.-B. Marchand {\it et~al.}, J. Cell Biol. {\bf 130},  331  (1995).

\bibitem{Argiro85}
V. Argiro, M. Bunge, and J. Johnson, J. Neurosci. Res. {\bf 13},  149  (1985).

\bibitem{Felder90}
S. Felder and E.~L. Elson, J. Cell Biol. {\bf 111},  2513  (1990).

\bibitem{Tilney89}
L.~G. Tilney and D.~A. Portnoy, J. Cell Biol. {\bf 109},  1597  (1989).

\bibitem{Sechi97}
A.~S. Sechi, J. Wehland, and J.~V. Small, J. Cell Biol. {\bf 137},  155
  (1997).

\bibitem{Cameron99}
L.~A. Cameron, M.~J. Footer, A. van Oudenaarden, and J.~A. Theriot, Proc. Natl.
  Acad. Sci. {\bf 96},  4908  (1999).

\bibitem{Abraham99}
V.~C. Abraham, V. Krisnamurthi, D.~L. Taylor, and F. Lanni, Biophys. J. {\bf
  77},  1721  (1999).

\bibitem{Rotsch99}
C. Rotsch, K. Jacobson, and M. Radmacher, Proc. Natl. Acad. Sci. USA {\bf 96},
  921  (1999).

\bibitem{Dogterom97}
M. Dogterom and B. Yurke, Science {\bf 278},  856  (1997).

\bibitem{Peskin93}
C.~S. Peskin, G.~M. Odell, and G.~F. Oster, Biophysical Journal {\bf 65},  316
  (1993).

\bibitem{Hill87}
T.~L. Hill, {\em Linear Aggregation Theory in Cell Biology} (Springer-Verlag,
  New York, 1987), Chap.~2.

\bibitem{Mogilner96}
A. Mogilner and G. Oster, Biophys. J. {\bf 71},  3030  (1996).

\bibitem{Mogilner96a}
A. Mogilner and G. Oster, Eur. Biophys. J. {\bf 25},  47  (1996).

\bibitem{Mogilner99}
A. Mogilner and G. Oster, Eur. Biophys. J. {\bf 28},  235  (1999).

\bibitem{Doi98}
M. Doi and S.~F. Edwards, {\em The Theory of Polymer Dynamics} (Clarendon
  Press, Oxford, 1998), Chap.~3.

\bibitem{Note1}
In order to avoid an excessive number of excursions back and forth across the
  circle radius, the added monomer is placed a small distance inside R.

\bibitem{Note2}
In two dimensions, the time for a particle to diffuse to capture is not
  strictly proportional to the area of the region in which it diffuses, but
  contains logarithmic corrections. Therefore the calculated velocities are not
  strictly independent of $R$. I have verified by use of a few test cases with
  larger values of $R$ that the predicted logarithmic scaling is observed.
  Extrapolating to a value of $R$ corresponding to a physiological
  interparticle spacing for actin monomers would modify the calculated
  velocities by a constant factor of about two, which we have not included
  because our focus is the force-dependence of the velocity rather than its
  absolute magnitude.

\bibitem{Ouden99}
A. van Oudenaarden and J.~A. Theriot, Nature Cell. Biol. {\bf 1},  493  (1999).

\end{thebibliography}

\end{document}